\documentclass[12pt]{iopart}
\usepackage{amsfonts}
\begin{document}
\title{Fermi-transported spinor and Dirac equation in general relativity}
\author{Roman  Plyatsko}
\address{ Pidstryhach Institute of Applied Problems in Mechanics and
Mathematics\\ Ukrainian National Academy of Sciences, 3-b Naukova Str.,\\ Lviv, 79060,
Ukraine}

\ead{plyatsko@lms.lviv.ua}

\begin{abstract}

The Fermi transport of the Dirac spinor is considered as a
generalization of the parallel transport of this spinor which was
introduced by V.~Fock and D.~Ivanenko (1929). The possible
structure of the new variant of the general-relativistic Dirac
equation based on the Fermi transport is discussed.

\end{abstract}

\pacs{ 0420, 9530S}

\maketitle

The generalization of the usual Dirac equation on the case of
general relativity was obtained in 1929 \cite{1, 2, 3}. The
introduction of the parallel transport of the Dirac spinor was the
main point in the procedure of derivation of the general-covariant
Dirac equation.

The equations describing the motion of a macroscopic spinning test particle in general
relativity were obtained in 1937 \cite{4}. According to these equations, the 4-vector of
spin $s^\mu$ undergoes to the Fermi transport \cite{5}
\begin{equation}\label{1}
\frac{Ds^\mu}{ds}=u^\mu\frac{Du_\sigma}{ds}s^\sigma
\end{equation}
where $s$ and $u^\mu$ are the proper time and the 4-velocity of a
spinning particle, $D/ds$ is the covariant derivative. In the case
when the world line of a particle is the geodesic line, i.e.
$Du_\sigma/ds=0$, we have from (\ref{1}) $Ds^\mu/ds=0,$ that is
the 4-vector of spin is parallel-transported. However, in general
for a spinning particle $Du_\sigma/ds\ne 0$ \cite{4}. Therefore,
in the context of investigations from \cite{1, 2, 3}, it is
interesting to consider the Fermi transport for the Dirac spinor
and to obtain the general-covariant Dirac equation which is based
just on the Fermi transport, in contrast to the equation from
\cite{2, 3}.

Here, in our preliminary estimations, we shall follow \cite{3},
including the corresponding notations. The variation of the tetrad
components of the 4-vector $A_i$ which is parallel-transported is
written in \cite{3} as
\begin{equation}\label{2}
\delta A_i=e_ke_l\gamma_{ikl}A_kdx_l
\end{equation}
where $e_1=e_2=e_3=-1, e_0=+1$ (i.e. the signature of the
Monkowski metric is $+---$), $\gamma_{ikl}$ are the Ricci
coefficients of rotation by definition
\begin{equation}\label{3}
\gamma_{ikl}=h_{\beta k}h^{\sigma}_lD_{\sigma}h^\beta_i,
\end{equation}
$D_\sigma$ is the covariant derivative by $x^\sigma$, $h^\sigma_l$
are the tetrad components. Here and in the following, the latin
indices are used for the local (tetrad) quantities and the greek
indices correspond to the global values.

Let us write the variation of the 4-vector $A_i$ in the case of
the Fermi transport. Analogously to (1), in the global coordinates
we have
\begin{equation}\label{4}
\frac{DA^\mu}{ds}=u^\mu\frac{Du_\sigma}{ds}A^\sigma.
\end{equation}
Using the expressions
$$
\frac{DA^\mu}{ds}=\frac{dA^\mu}{ds}+\Gamma^\mu_{\nu\lambda}A^\nu
u^\lambda,
$$
\begin{equation}\label{5}
\frac{dA^\mu}{ds}=\frac{dh^\mu_i}{ds}A^i+h^\mu_i\frac{dA^i}{ds}
\end{equation}
we obtain from (4) the relationship in the local components
\begin{equation}\label{6}
\frac{dA^i}{ds}=-h^i_\mu \frac{dh^\mu_k}{ds}A^k-h^i_\mu
\Gamma^\mu_{\nu\lambda}h^\nu_k A^k
u^\lambda+u^k\frac{Du_l}{ds}A^l.
\end{equation}
Taking into account the known connection between
$\Gamma^\mu_{\nu\lambda}$ and $\gamma^\mu_{\nu\lambda}$
\begin{equation}\label{7}
 \Gamma^\mu_{\nu\lambda}=\gamma^\mu_{\nu\lambda}+h^\mu_k
 h^k_{\nu,\lambda}
\end{equation}
according to (6) we write
\begin{equation}\label{8}
\delta A^i=(-\gamma^i_{kl}+\delta^i_l\frac{Du_k}{ds}) A^k dx^l.
\end{equation}
where $\delta^i_l$ is the Kroneker symbol. In terms of eq. (2) it
follows from (8)
\begin{equation}\label{9}
\delta A_i=e_k e_l(\gamma_{ikl}+\delta_{il}a_k)A_k dx_l
\end{equation}
where we denote
\begin{equation}\label{10}
a_k=\frac{Du_k}{ds}
\end{equation}
and $\delta_{il}$ is connected with $\delta^l_i$ by the relation
$\delta^l_i=e_l\delta_{li}$.

So, expression (9) for the Fermi transport of the vector $A_i$
differs from the corresponding expression for the parallel
transport by the second term in the brackets of eq. (9) containing
the 4-vector $a_k$.

Let us consider the variation of the spinor components at the
Fermi transport using the procedure described in \cite{3}, eqs.
(13)-(18), for the parallel transport of spinor. According to eq.
(13) from \cite{3} we write
\begin{equation}\label{11}
\delta\psi=e_i C_i\psi dx_i
\end{equation}
where $\psi$ is the 4-component spinor, $C_i$ are the matrices.
Using the equations analogous to eqs. (14)-(17) from \cite{3}, we
get
\begin{equation}\label{12}
C_l=\frac{1}{4}\alpha_m\alpha_k e_k\gamma_{mlk}+i\Phi_l+a_l
\end{equation}
where $\alpha_m$ are the Dirac matrices and $\Phi_l$ are the
matrices described in \cite{3} after eq. (18). So, expression (12)
differs from eq. (18) in \cite{3} by the term $a_l$.

Probably, to obtain the Dirac equation which is based on the Fermi
transport of the spinor $\psi$ (11), (12), it is possible to
follow eqs. (19)-(23) from \cite{3}.

For the matrices $\Gamma_\sigma$ which are determined as
\begin{equation}\label{13}
\Gamma_\sigma=e_k h_{\sigma k}C_k
\end{equation}
(see eq. (20) in \cite{3}) according to (12) we get
\begin{equation}\label{14}
\Gamma_\alpha^+\gamma^\sigma +
\gamma^\sigma\Gamma_\alpha=-D_\alpha\gamma^\sigma+\delta^\sigma_\alpha
e_m\alpha_m a_m
\end{equation}
where by definition
\begin{equation}\label{15}
\gamma^\sigma=e_k\alpha_k h^\sigma_k.
\end{equation}
Expression (14) is the generalization of expression (24) from
\cite{3}
\begin{equation}\label{16}
\Gamma_\alpha^+\gamma^\sigma +
\gamma^\sigma\Gamma_\alpha=-D_\alpha\gamma^\sigma,
\end{equation}
which was obtained for the parallel transport of the spinor, on
the case of the Fermi transport.

We stress that just relationship (16) is important for the
conclusion that the general-covariant Dirac equation can be
written as \cite{3}
\begin{equation}\label{17}
\frac{h}{2\pi i}\gamma^\sigma(\frac{\partial \psi}{\partial
x^\sigma}-\Gamma_\sigma\psi)-mc\alpha_4\psi=0.
\end{equation}
In particular, due to (16) it follows from (17) that the
divergation of the 4-vector of the current density
\begin{equation}
J^\rho=\psi^+\gamma^\rho\psi
\end{equation}
is equal to 0:
\begin{equation}\label{19}
\frac{1}{\sqrt{g}}\frac{\partial }{\partial
x^\sigma}(\sqrt{g}J^\sigma)=0.
\end{equation}

The simple generalization of eq. (17) on the case of the Fermi
transport is
\begin{equation}\label{20}
\frac{h}{2\pi i}\gamma^\sigma(\frac{\partial \psi}{\partial
x^\sigma}-\Gamma_\sigma\psi-e_k h_{\sigma k}a_k \psi
)-mc\alpha_4\psi=0
\end{equation}
where $\Gamma_\sigma$ are the same matrices that are present in
(17), i.e. calculated in the case of the parallel transport. The
term in (20) which is proportional to $a_k$ can be easily obtained
after substituting expression (12) in (13). Naturally, the
4-vector $a_k$ in (20) must be expressed through the quantum
values, in particular through $\psi^+$ and $\psi$. It is necessary
to carry out more detailed investigations to determine the
explicit form of the 4-vector $a_k$ and we shall consider this
problem in another paper. Here we point out only that from eqs.
(17), (20) follows eq. (19) in the case when $a_k$ satisfies the
relationships
\begin{equation}\label{21}
e_m\psi^+\alpha_m a_m\psi=0,\qquad a_\sigma J^\sigma =0.
\end{equation}
Probably, eqs. (21) shall be useful for the determination of the
structure of $a_m$.

\end{document}